\documentclass[10pt]{iopart}

\usepackage[T1]{fontenc}
\usepackage[utf8]{inputenc}
\usepackage{graphicx}
\usepackage{tabularx}
\usepackage{color, soul}
\usepackage{textcomp}
\usepackage{enumerate}
\usepackage{todonotes}
\usepackage[draft]{hyperref}
\usepackage{comment}
\usepackage{subcaption}
\usepackage{url}
\usepackage[normalem]{ulem}
\usepackage{makecell}
\usepackage{stfloats}
\usepackage{booktabs} 
\usepackage{multirow}
\usepackage{cite}

\begin{document}

\title[Investigation of Plasma Mixing Processes in the Context of Indirect Drive Inertial Confinement Fusion]{Investigation of Plasma Mixing Processes in the Context of Indirect Drive Inertial Confinement Fusion} 

\author{Xiaoran Li$^{1}$, Jie Qiu$^{1}$, Shuqing Zhang$^{1}$, Liang Hao$^{1}$\footnote{Corresponding author: hao\_liang@iapcm.ac.cn}, Shiyang Zou$^{1}$}

\address{
$^1$ Institute of Applied Physics and Computational Mathematics, Beijing 100094, People’s Republic of China}

\vspace{10pt}
\begin{indented}
        \item[]\today
\end{indented}

\begin{abstract}
In inertial confinement fusion (ICF), the dynamics of plasma mixing in hohlraums critically influence laser-plasma instabilities (LPI) and implosion performance. This study investigates the mixing of hohlraum ablated Au plasmas and filling C$_5$H$_{12}$ plasmas using one-dimensional particle-in-cell (PIC) simulations.
We find that ion-ion collisions slow the diffusion of ions, rendering Au ions sub-diffusive, while C and H ions remain super-diffusive. Due to their lower collisionality, H ions diffuse faster into Au regions than C ions, leading to a distinct separation between C and H ions at the interface. Although an electrostatic shock is still generated at the plasma interface in the presence of collisions, its electric field strength and propagation speed are notably reduced.
To systematically explore plasma mixing in hohlraum environments, we evaluate the individual effects of incident laser irradiation, plasma flow, and inhomogeneous density profiles on ion mixing. We find that laser irradiation and plasma flow have a minor impact on ion mixing compared to diffusion-driven processes, while the inhomogeneous density profile restricts diffusion from low-density to high-density regions.
By incorporating realistic hohlraum plasma conditions derived from radiation hydrodynamic models into the PIC simulations, we demonstrate that the diffusion of C and H ions continues to dominate ion mixing. Simple phenomenological fits are derived to describe the evolution of the mixing width in a hohlraum condition.
Further theoretical calculations indicate that the penetration of H and C into Au plasmas suppresses stimulated Brillouin scattering (SBS) within the mixing layer. This finding underscores the importance of integrating ion mixing effects into LPI codes for more accurate modeling of ICF hohlraum dynamics.

\end{abstract}

\ioptwocol

\section{Introduction}
\label{sec:intro}

In indirect inertial confinement fusion (ICF), ICF hohlraums are filled with various types of plasmas, including low-Z plasmas from filling gases, low-Z and mid-Z plasmas from the capsule's ablated material, and high-Z plasmas from the hohlraum wall. The dynamics of mixing between different plasmas can significantly affect laser plasma instabilities (LPI) and implosion performance~\cite{yao2020, meng2016b, zhang2024}. On one hand, the electrostatic field arising from the interaction of the expanding high-Z plasma from the inner wall and the ablated plasmas from the capsule can accelerate high-energy ions back toward the capsule, leading to implosion asymmetry~\cite{zhang2024a, liang2024, shan2018a, zhang2017}. On the other hand, the mixing of high-Z plasmas from hohlraum walls with filling low-Z plasmas can create complex plasma conditions in the hohlraum, which are difficult to accurately diagnose or simulate with current radiation hydrodynamic (RH) models~\cite{li2013b, shestakov2000}. This, in turn, can lead to inaccuracies in LPI calculations~\cite{yan2019, hao2019}. Therefore, understanding the dynamics of plasma mixing in ICF hohlraums is of significant importance for achieving successful ignition.

The mixing of plasmas has long been considered an important fundamental physics problem. Several fluid models have been developed to address the diffusion~\cite{amendt2010, amendt2011, kagan2012} and mixing~\cite{molvig2014c, molvig2014,yan2021} in plasmas with multi-ion species. However, hydrodynamic simulations cannot include kinetic effects, which can significantly impact plasma mixing behaviors~\cite{yan2019, yin2016, zhang2024a}. For example, strong shock waves generated at the interfaces between high-temperature Au bubbles and relatively low-temperature deuterium plasmas can reflect and accelerate deuterium ions~\cite{zhang2024a, liang2024}. Even at a quasi-pressure-balanced high-Z/low-Z plasma interface, ``super-diffusive'' behavior is still observed~\cite{yan2019, yin2016}. Yan \textit{et al.}~\cite{yan2019} found that a diffusion-driven collisionless shock wave was generated at an initially sharp Au-He plasma interface, with the same plasma density and temperature on both sides. This collisionless shock wave leads to a mixing width that grows as $w_m \propto t^\alpha$, where $\alpha \approx 1$, which is significantly faster than the classical mixing suggested by fluid theory. Although Yan \textit{et al.} suggested that collisional effects were negligible in their 2\,ps study~\cite{yan2019}, questions remain about how this electrostatic shock wave evolves with collisions over longer timescales, and how plasma mixing evolves under the combined influence of both the shock wave and collisions.

In this paper, we investigate the mixing between blow-off high-Z plasmas and filling low-Z plasmas in ICF hohlraums using 1D particle-in-cell (PIC) simulations. We first extend the work of~\cite{yan2019} by examining the diffusion-driven plasma mixing over a longer timescale, allowing the effects of collisions to manifest. Next, to better understand plasma mixing in the complex hohlraum environment, we systematically investigate the individual effects of incident laser, plasma flow, and inhomogeneous density profiles on ion mixing. Finally, PIC simulations with realistic hohlraum plasma parameters derived from RH calculations are performed, and the ion mixing behaviors and their potential effects on LPIs are discussed.


\section{Simulation setup and basic parameters}
\label{sec:setting}

To investigate the evolution of multi-ion species, we perform a series of one-space-dimensional and three-velocity-dimensional (1D3V) PIC simulations using the open-source code EPOCH~\cite{arber2015}.
We investigate the mix between hohlraum ablated Au plasmas and C$_5$H$_{12}$ plasmas from filling gases~\cite{hao2019,zhao2019a}.
To better understand the ion-mixing behavior under real hohlraum conditions, and to reduce computational costs, we use two different simulation domains in this study.

The first is a relatively short domain of 200$\lambda_0$ ($\lambda_0 = 0.351\,\mu$m) with simplified plasma conditions, labeled as domain A. This domain is used to study the effects of individual mechanisms on ion-mixing dynamics, such as collisions, laser radiation, plasma flow, and density gradients.
The left half contains C$_5$H$_{12}$ plasmas, and the right half is filled with Au plasmas. The plasma density is uniform across the entire domain, with a density of $n_e = 0.1 n_c$, unless otherwise indicated. Here, $n_c = 8.9 \times 10^{21}$\,cm$^{-3}$ is the critical density for a laser with a wavelength of $0.351\,\mu$m. The 200$\lambda_0$ plasma domain is discretized into 6400 grids, with a grid size of $\lambda_0$/32. The temperatures of electrons and ions are also uniform across the entire domain, with an electron temperature of $T_e = 3000$\,eV and an ion temperature of $T_i = 1000$\,eV.

\begin{figure*}[hbt]
	\centering
	\includegraphics[width=0.32\textwidth]{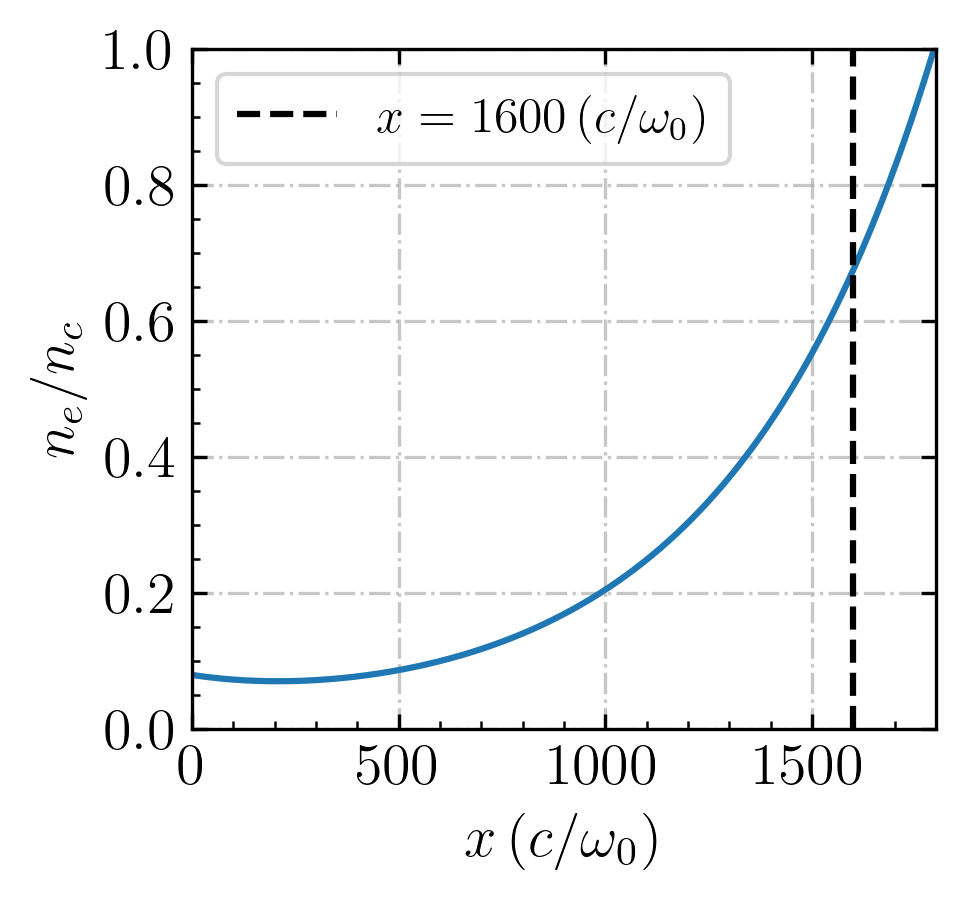}
	\includegraphics[width=0.32\textwidth]{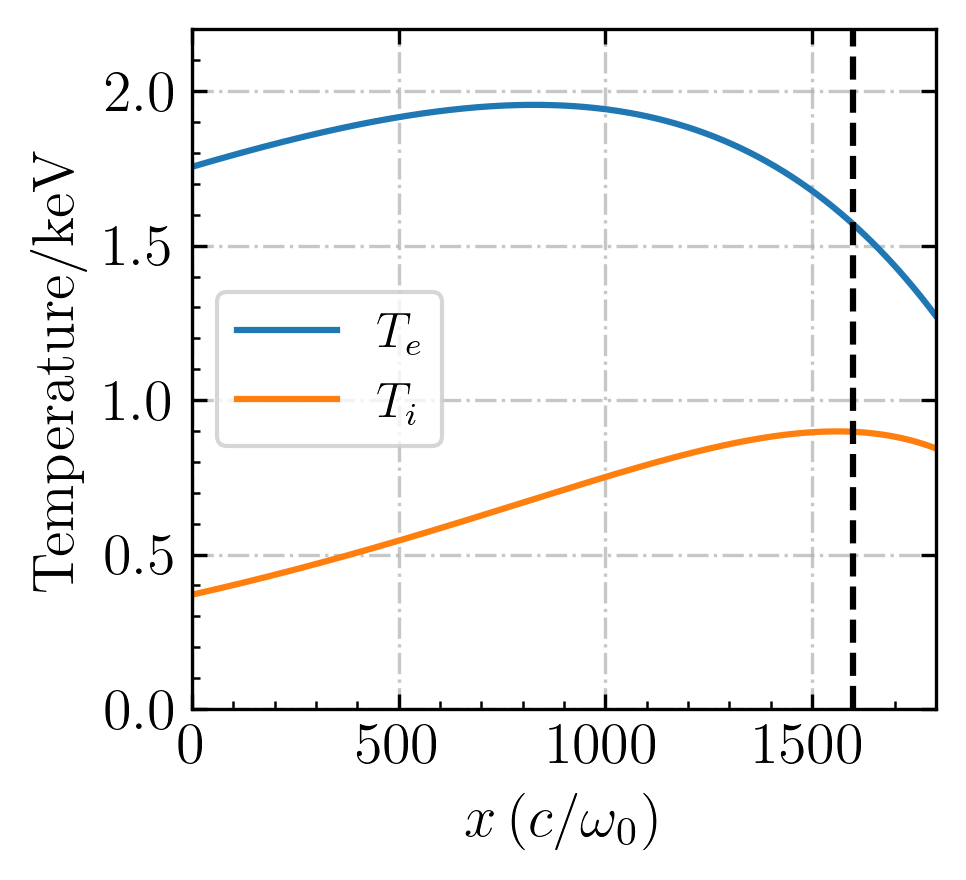}
	\includegraphics[width=0.32\textwidth]{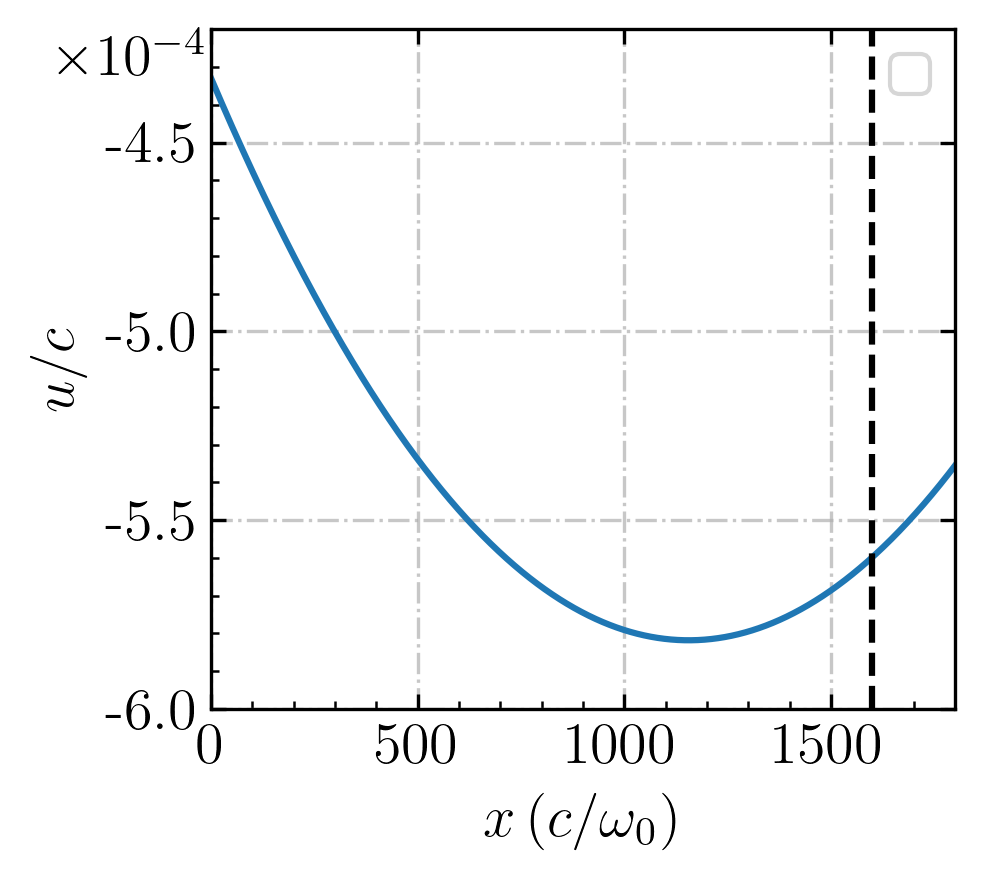}
	\caption{Plasma parameters extracted from RH simulations at 1.1\,ns in~\cite{hao2019}, which is used as the initial plasma state for domain B.}
	\label{fig:initial_condition}
\end{figure*}

The second, labeled as domain B, is a 100$\mu$m simulation domain with plasma parameters extracted from RH simulations, used to analyze complex ion behaviors in a real ICF hohlraum. The plasma conditions in the hohlraum are complex. The high-Z Au is significantly heated by the incident laser and expands into the filling gases~\cite{zhang2025}, leading to nonuniform density and temperature distributions for both electrons and ions, as well as a background plasma flow.
We use simulation conditions based on an experiment conducted at the SG-100kJ laser facility, with plasma parameters extracted from RH simulations at 1.1\,ns in~\cite{hao2019}, as shown in figure~\ref{fig:initial_condition}.
The domain length is 1800$c/\omega_0$ ($\approx$ 100\,$\mu$m), where $c$ is the speed of light and $\omega_0$ is the frequency of the laser with a wavelength of $\lambda_0$. The left 1600$c/\omega_0$ contains C$_5$H$_{12}$ plasmas, and the right 200$c/\omega_0$ contains Au plasmas. This domain is discretized into 9000 grids, with a grid size of 0.2$c/\omega_0$.

 
Besides the focused plasma domain, we have two convolutional perfectly matched layers~\cite{roden2000} of 80 grids at both edges, and thermal boundaries are applied for particles. The ion densities are set according to the electron density and the charge number of the ions. Specifically, we have $n_\mathrm{Au} = n_e / 50$, $n_\mathrm{C} = 5 n_e / 42$, and $n_\mathrm{H} = 2 n_e / 7$. The simulations use 1000 computational macro-particles per cell for electrons and Au ions, and 500 macro-particles per cell for C and H ions. Binary collisions between particles~\cite{nanbu1998} are considered, including electron-electron, electron-ion, and various ion-ion collisions. We simulate the evolution of all particles for at least 20\,ps. The time step is adaptively chosen according to the CFL condition, averaging approximately 0.03\,fs.

%


\section{Simulation results}

\subsection{Effect of collisions}
\label{sec:col}

To investigate the effects of collisions on ion diffusion behaviors, we run five simulation cases using the simulation domain A described in section~\ref{sec:setting}. Case a excludes all types of collisions; case b includes all types of collisions; case c contains only ion-ion collisions; case d contains only electron-ion collisions; and case e contains only electron-electron collisions. By comparing cases a and b, we find that the diffusion velocity of all three ions is limited by collisions.

\begin{figure*}[htb]
	\centering
	\includegraphics[width=0.495\textwidth]{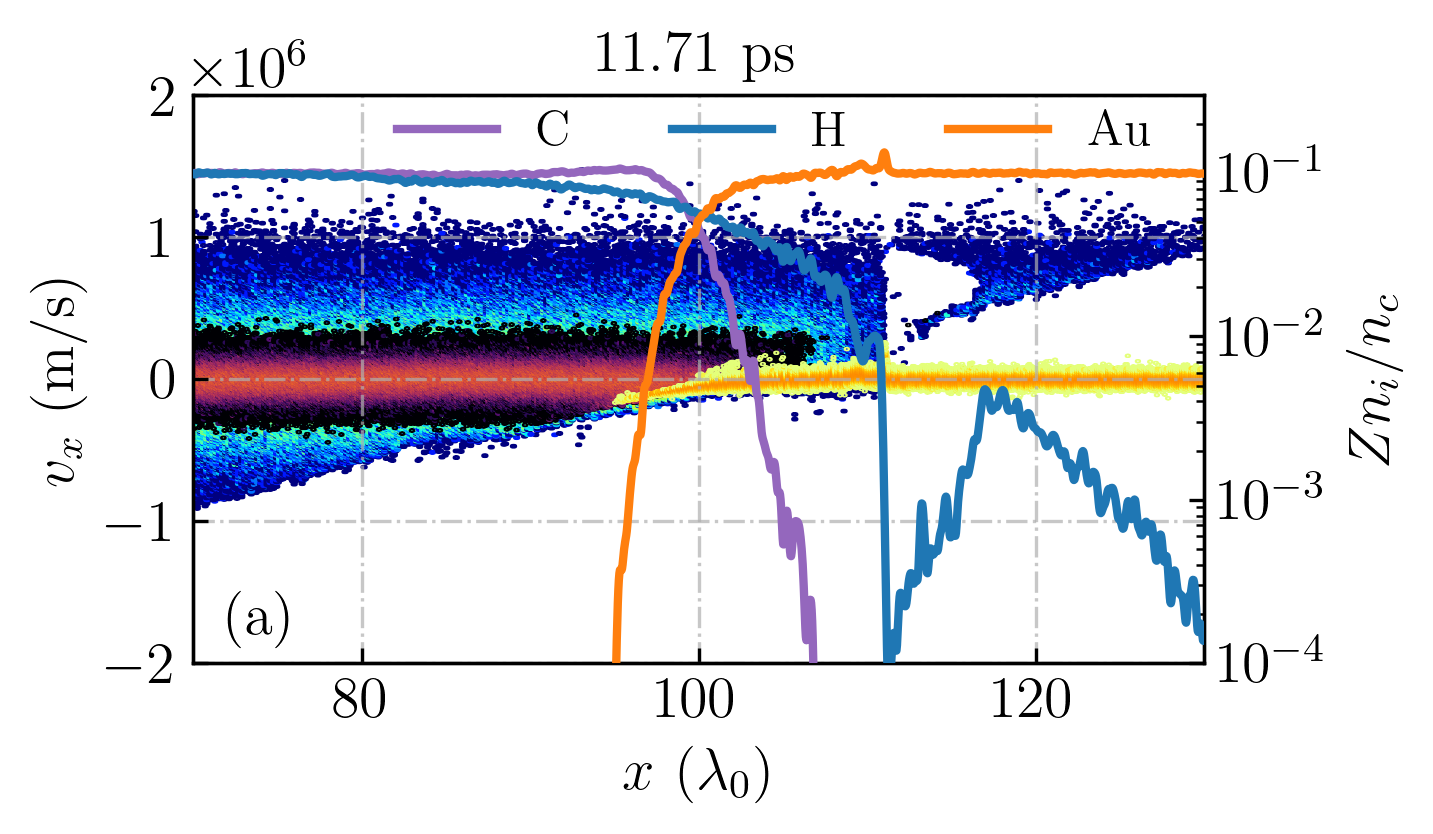}
	\includegraphics[width=0.495\textwidth]{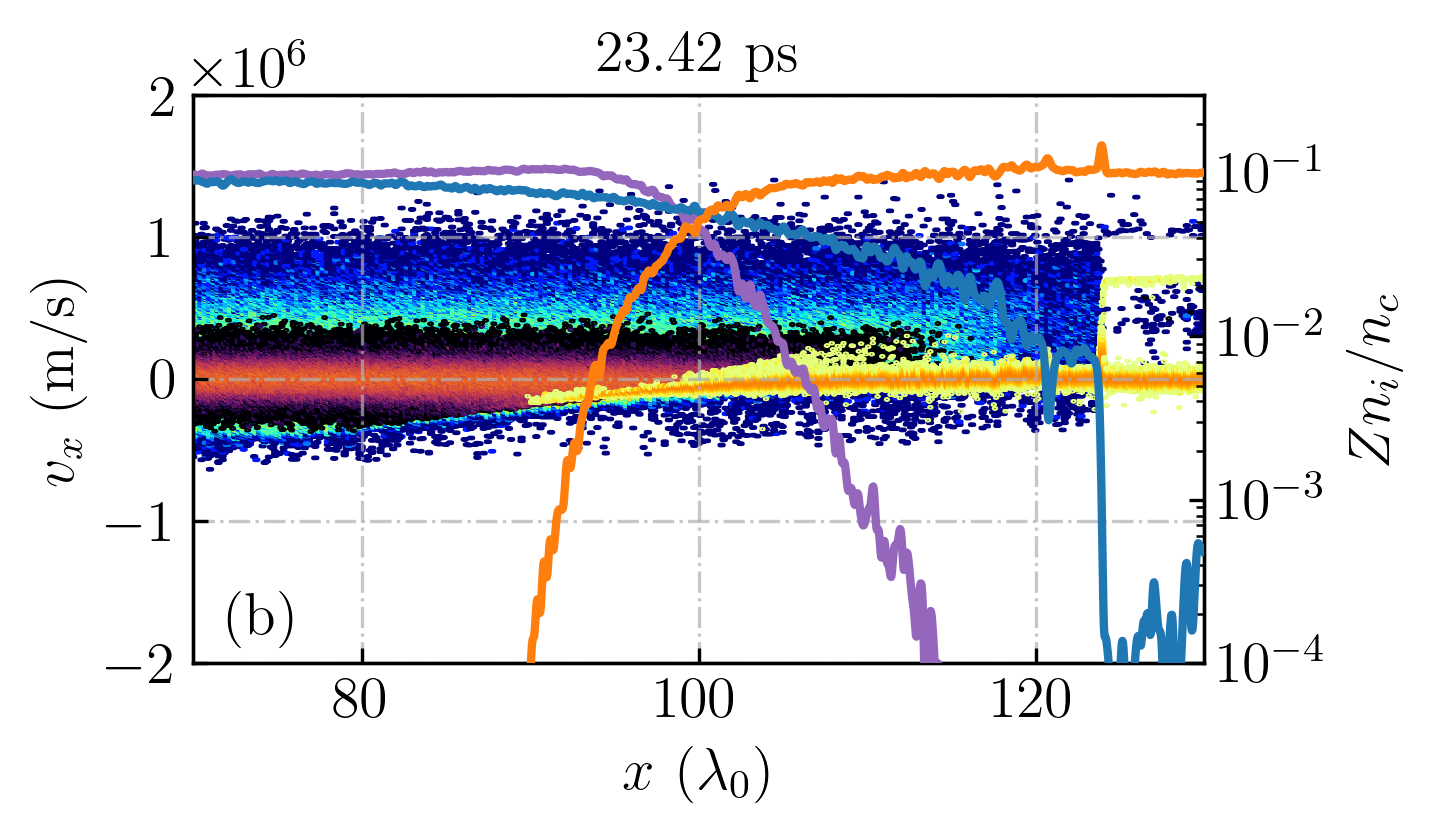}
	\includegraphics[width=0.495\textwidth]{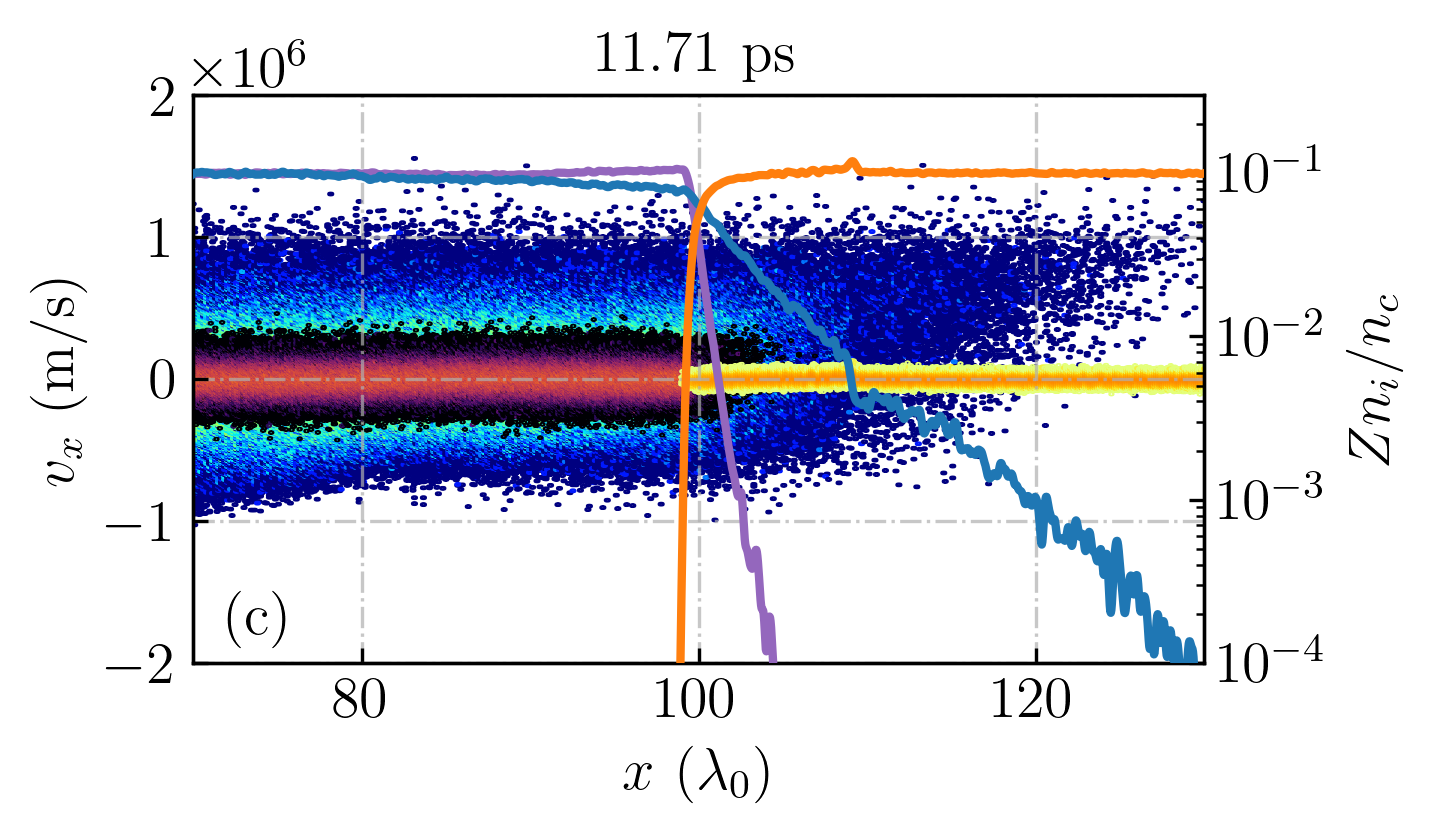}
	\includegraphics[width=0.495\textwidth]{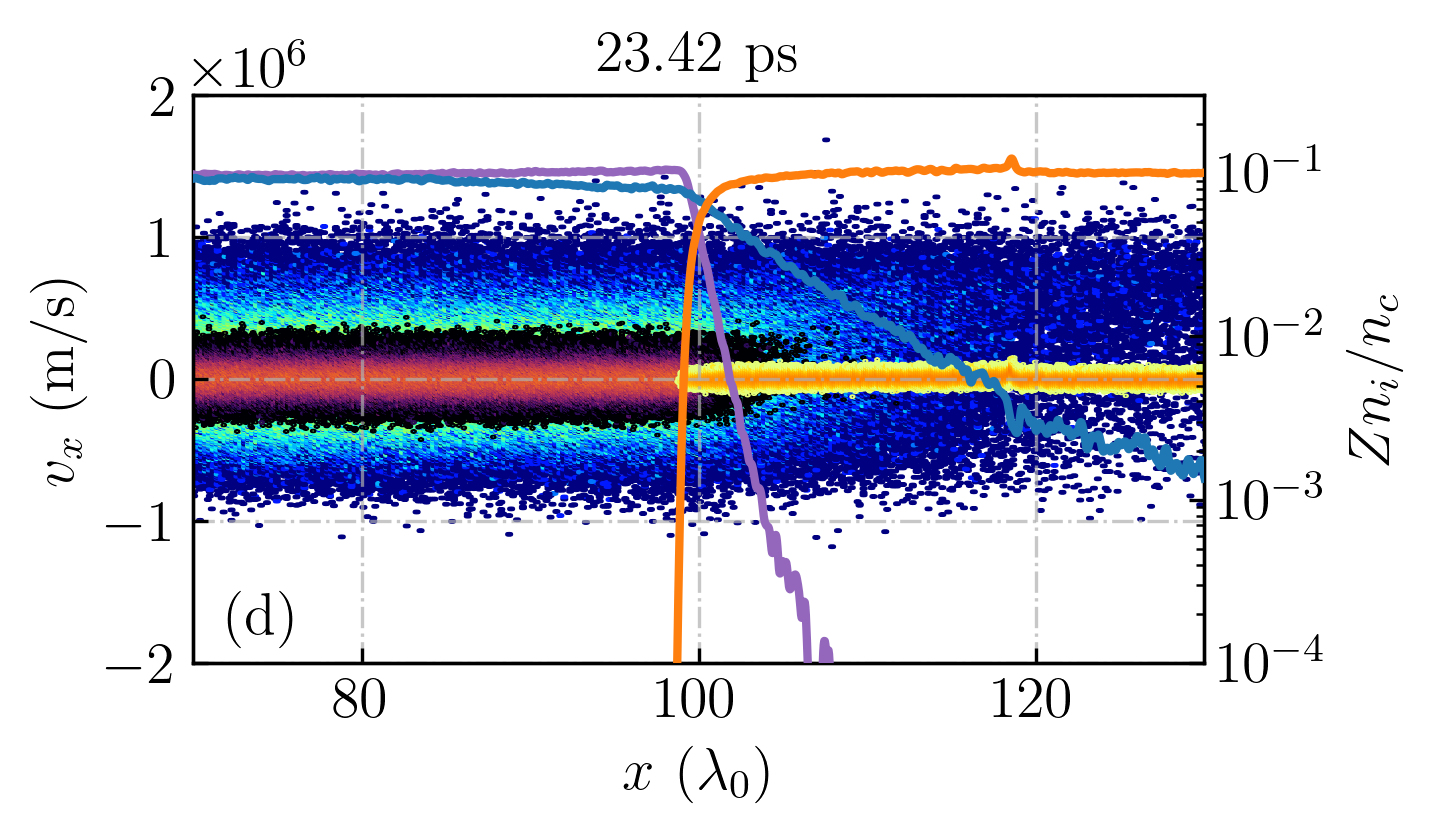}
	\caption{The phase-space distribution plots $v_x$-$x$ of the C and H ions (left half) and Au ions (right half), the ions density of C, H and Au ions in logarithm scale. The quantities are plotted at different cases: (a) collisionless case at 11.71\,ps , (b) collisionless case at 23.42\,ps , (c) collisional case at 11.71\,ps and (d) collisional case at 23.42\,ps.}
	\label{fig:phase_den_col_wocol}
\end{figure*}

\begin{figure*}[htb]
	\centering
	\includegraphics[width=0.495\textwidth]{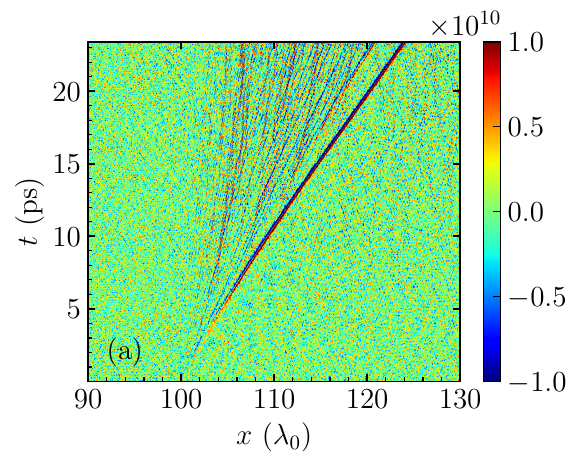}
	\includegraphics[width=0.495\textwidth]{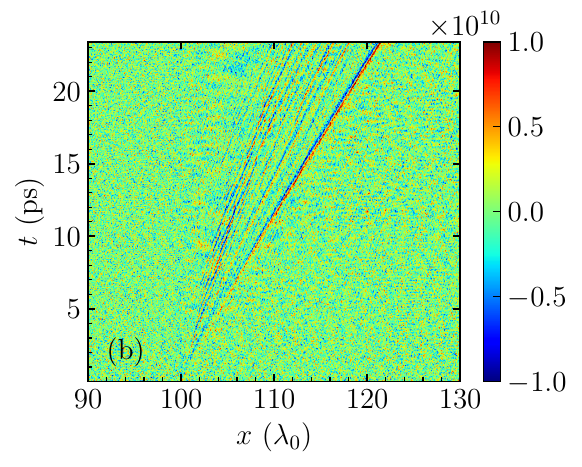}
	\caption{The temporal evolution of the electric field in the $x$ direction from 0 to 23.42\,ps. Subfigures (a) and (b) are for the collisionless case a and the collisional b, respectively.}
	\label{fig:ex_x_t}
\end{figure*}

Figure~\ref{fig:phase_den_col_wocol} shows the phase-space distribution of C, H, and Au ions for cases a and b. The mixing of Au-C and Au-H ions near the interface can be clearly seen in both cases. In the case without collisions, the mixing of Au and H ions is somewhat similar to the mixing of Au-He ions in~\cite{yan2019}, where ions expand into the other medium in the form of a rarefaction wave, and an electrostatic shock wave occurs at the interface, as shown in figure~\ref{fig:ex_x_t}.
In figure~\ref{fig:ex_x_t}(a), the precursor H ions in front of the electrostatic shock wave are accelerated by this potential hump, and those H ions with lower energy behind the shock wave are blocked by the potential hump, thus a hole is formed in the phase-space distribution of H ions near the potential hump, as shown in figure~\ref{fig:phase_den_col_wocol}(a) and (b).
Au ions in front of the wave which tends to diffuse to the left are also blocked and even reflected by the shock wave due to the strong field, as shown in figure~\ref{fig:ex_x_t}(b), and the Au ion density forms a peak at the potential hump position.
For the collisional case b, the electrostatic shock wave still occurs at the interface and causes a bend in the phase-space distribution of Au ions.
However, the amplitude of the electrostatic field is much smaller than in the collisionless case, as can be clearly seen in figure~\ref{fig:ex_x_t}.
The hole in the H ions phase-space distribution and the reflections of Au ions disappear due to the reduced electrostatic field, as shown in figure~\ref{fig:phase_den_col_wocol}(c) and (d).
Additionally, the ion diffusion distances are significantly limited for the collisional case, especially for Au ions.
Furthermore, an electron density peak propagates to the right together with the shock wave. As the collisional effects weaken the electrostatic wave amplitude, the peak electron density also decreases, and the velocity of the electron density peak decreases with the shock wave, from 1.2$c_{s,\mathrm{Au}}$ to $c_{s,\mathrm{Au}}$, where $c_{s,\mathrm{Au}}$ is the ion-acoustic speed of Au ions.
By comparing cases a-e, we found that electron-ion collisions are responsible for slowing the electrostatic shock wave by reducing the charge separation.

\begin{figure}[htb]
	\centering
	\includegraphics[width=0.49\textwidth]{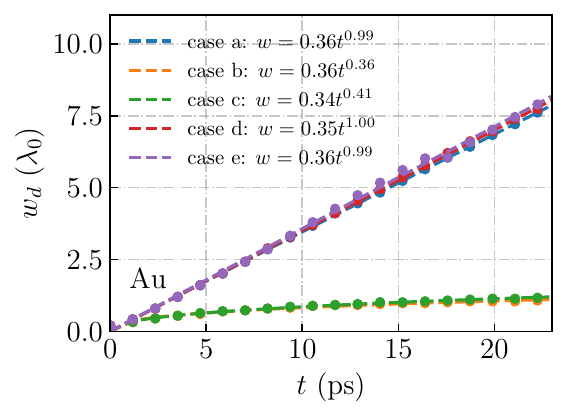}
	\includegraphics[width=0.49\textwidth]{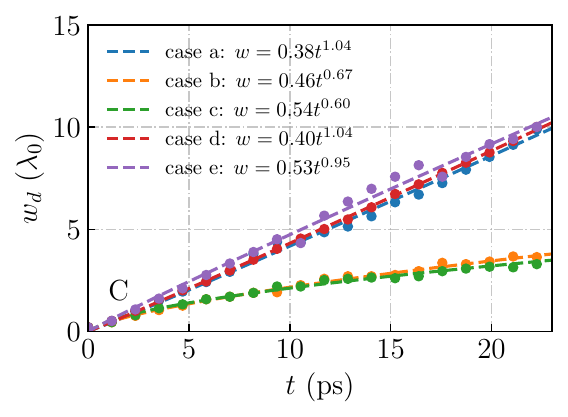}
	\includegraphics[width=0.49\textwidth]{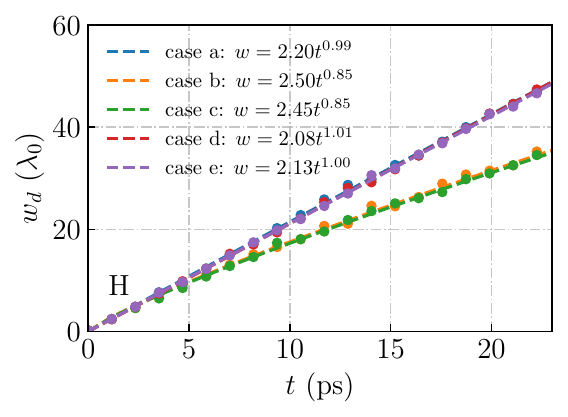}
	\caption{The diffusion width of different ions is measured as a function of time for collisionless case a and cases b-e with different types of collisions, where case b contains all types of collisions, case c contains only ion-ion collisions, case d contains only electron-ion collisions, case e contains only electron-electron collisions.}
	\label{fig:wd_t_diffcol}
\end{figure}

To quantitatively analyze the diffusion velocity for different cases, we plot the diffusion width ($w_d$) versus time for Au, C, and H ions in figure~\ref{fig:wd_t_diffcol}, respectively. The diffusion width ($w_d$) is defined as the distance between the initial interface position (100$\lambda_0$) and the edge position of a species, where the edge position corresponds to the location where the species density is 0.001$n_c$/Z~\cite{yin2016}. It can be seen that $w_d$ is almost proportional to time for all three ions in cases a, d, and e, which is similar to the findings in~\cite{yan2019}. However, for cases b and c, where ion-ion collisions are considered, the diffusion velocity is significantly slower. We fit the diffusion width versus time as $w_d \propto t^\alpha$ for all three ions. For the collisional case b, $\alpha$ is 0.36, 0.67, and 0.85 for Au, C, and H ions, respectively. For C and H ions, $\alpha$ is larger than 0.5, a value suggested by classic fluid theory, indicating super-diffusive behavior. In contrast, Au ions, with $\alpha$ smaller than 0.5, exhibit sub-diffusive behavior~\cite{cecconi2022}.

Comparing our study with the mixing of He and Au plasmas in~\cite{yan2019}, the effect of collisions are clearly reveal here as we simulate the mixing process for a longer time.
Another notable point is that species separation between H and C is observed here due to 
different diffusion velocities.
This leads to another phenomena that the propagation speed of the electrostatic shock is between the diffusion velocity of C and H ions.
Since the diffusive C ions are behind the shock wave, the electrostatic shock wave is not responsible for accelerating the diffusion of C ions.
Therefore, we suggest that this super-diffusive transport for C ions is a manifestly kinetic behavior, resulting from the free streaming of light ions before the collisional effects fully dominate.
The precursor H ions in front of the shock wave can be accelerated by this wave, but the relatively lower collisional property of H ions would also be an important reason for the larger $\alpha$.
The Au ions exhibit sub-diffusive behavior ($\alpha$ < 0.5) due to frequent ion-ion collisions within themselves and with C and H ions.
As C and H ions move to the right with higher thermal velocities, they collide with Au ions, which hinders the diffusion of Au ions in the opposite direction.

\subsection{Effects of incident laser, plasma flow, and density gradient}
\label{sec:diffmec}

To better understand plasma mixing in the hohlraum environment in a relatively simple condition, we test the effects of incident laser, plasma flow, and density gradient separately in this section using four new cases f-i. The settings of cases f-i are based on the case b with all types of collisions described in section~\ref{sec:col}, which utilizes the 200$\lambda_0$ domain A described in section~\ref{sec:setting} and has a uniform plasma density of $n_e = 0.1 n_c$.
For case f, we attach an incident laser at the left boundary with an intensity of $2 \times 10^{15}$ W/cm$^2$. In case g, we apply a homogeneous flow with a velocity of $-1.5 \times 10^{5}$ m/s in the plasmas. Here, the negative sign indicates that the flow is directed to the left. In case h, we apply an inhomogeneous flow that linearly increases from $-1.8 \times 10^{5}$ m/s at the left boundary to $-1.2 \times 10^{5}$ m/s at the right boundary. For case i, we use an inhomogeneous plasma density where the electron density increases linearly from $n_e = 0$ at the left boundary to $n_e = 0.2 n_c$ at the right boundary.

\begin{figure}[htb]
	\centering
	\includegraphics[width=0.49\textwidth]{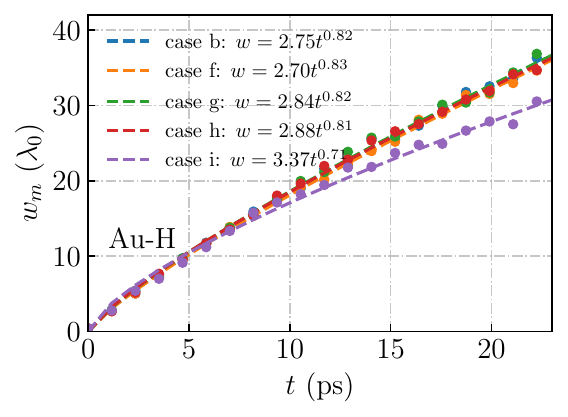}
	\includegraphics[width=0.49\textwidth]{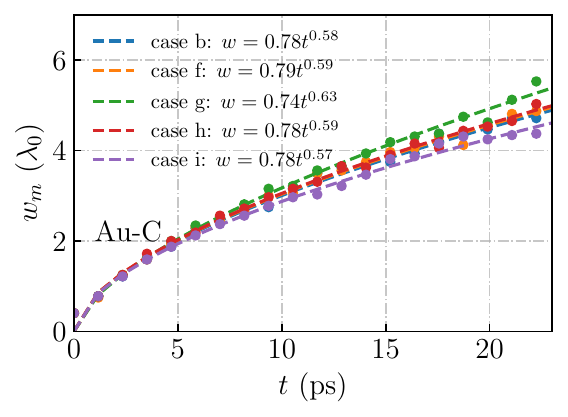}
	\caption{The mixing width of Au ions with C and H ions versus time for cases f-i, compared with case b in figure~\ref{fig:wd_t_diffcol}. Case f contains an incident laser at the left boundary, case g contains a homogeneous flow with a velocity of $-1.5 \times 10^{5}$ m/s, case h contains an inhomogeneous flow, and case i has an inhomogeneous plasma density.}
	\label{fig:wm_t_diffmec}
\end{figure}

Since the initial high-Z/low-Z interface moves together with the flow, we in this section use the mixing width ($w_m$) to characterize mixing behaviors between Au ions with C and H ions instead of the diffusion width ($w_d$) used in section~\ref{sec:col}. Here the mixing width is defined as the distance between the front edge of Au and C or H ions, where the edge position corresponds to the location where the species density is $0.001 n_c / Z$. 
Figure~\ref{fig:wm_t_diffmec} shows the mixing width between Au ions and C/H ions versus time for cases f-i, compared with case b.
The mixing speed between Au and H ions is almost unaffected by the incident laser and the plasma flow. However, it decreases when the inhomogeneous plasma density is introduced. This is because the density of the H ions is reduced on the left side of the interface compared to the standard case b, making it harder for H ions to diffuse from low-density side to high-density side. The lower density on the left side also slightly reduces the mixing between Au and C ions in a similar manner. 
The mixing speed between Au and C ions slightly increases when a plasma flow is applied in the leftward direction. This occurs because the interface moves to the left along with the flow, and collisions with Au ions trap some of the C ions, leading to a larger mixing width compared to case b at the same time.

\begin{figure}[htb]
	\centering
	\includegraphics[width=0.495\textwidth]{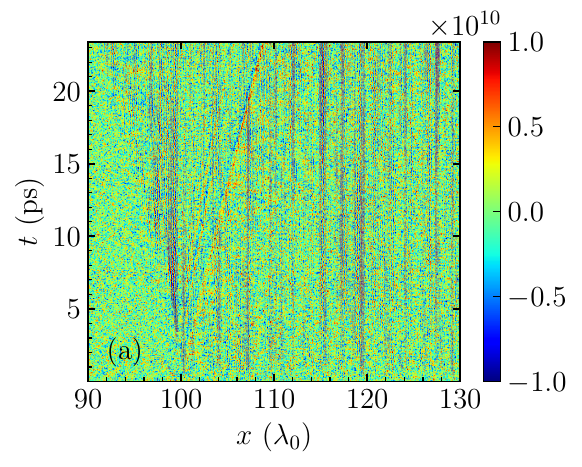}
	\includegraphics[width=0.495\textwidth]{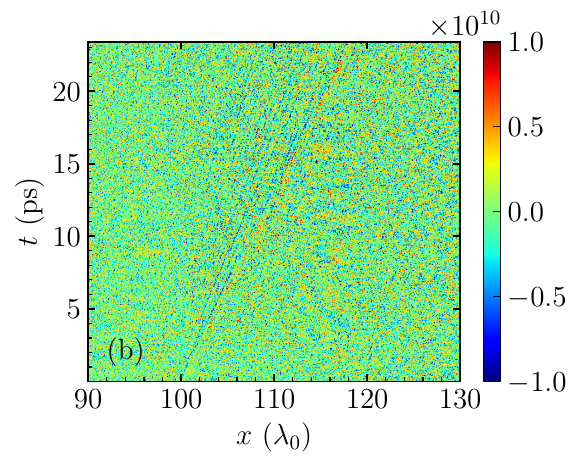}
	\caption{The temporal evolution of the electric field in the $x$ direction from 0 to 23.42\,ps. Subfigures (a) and (b) are for the case g with a flow of -1.5$\times$10$^{5}$\,m/s and the case i with an inhomogeneous plasma density, respectively.}
	\label{fig:ex_t_fl_ne}
\end{figure}

Though the plasma flow and inhomogeneous density have only a slight influence on the plasma mixing speed, they do impact the electrostatic shock wave, as shown in figure~\ref{fig:ex_t_fl_ne}. The speed of the electrostatic shock wave is significantly reduced with a flow velocity $v_\mathrm{fl} = -1.5 \times 10^5$\,m/s. The new shock wave velocity is $1.26 \times 10^5$\,m/s, which is equal to $c_{s,\mathrm{Au}} + v_\mathrm{fl}$. 
The flow also induces turbulence in the plasma, which generates additional noise in the longitudinal electric field. The inhomogeneous plasma density has little effect on the shock wave velocity, but it weakens the electrostatic field amplitude, making it less noticeable compared to the background noise in the field.
Combining figures~\ref{fig:ex_t_fl_ne} and~\ref{fig:wm_t_diffmec}, though the speed or the amplitude of the electrostatic shock wave has largely changed in case g and case i, the mixing speed between Au and C or H ions are still similar to each other, therefore we suggest that ion mixing, which is dominated by ion interactions, is less relevant to the electrostatic shock wave driven by electron-ion interactions.

\subsection{Evolution of ion-mix in ICF hohlraums}
\label{sec:hohlraum}

\begin{figure*}[htb]
	\centering
	\includegraphics[width=0.495\textwidth]{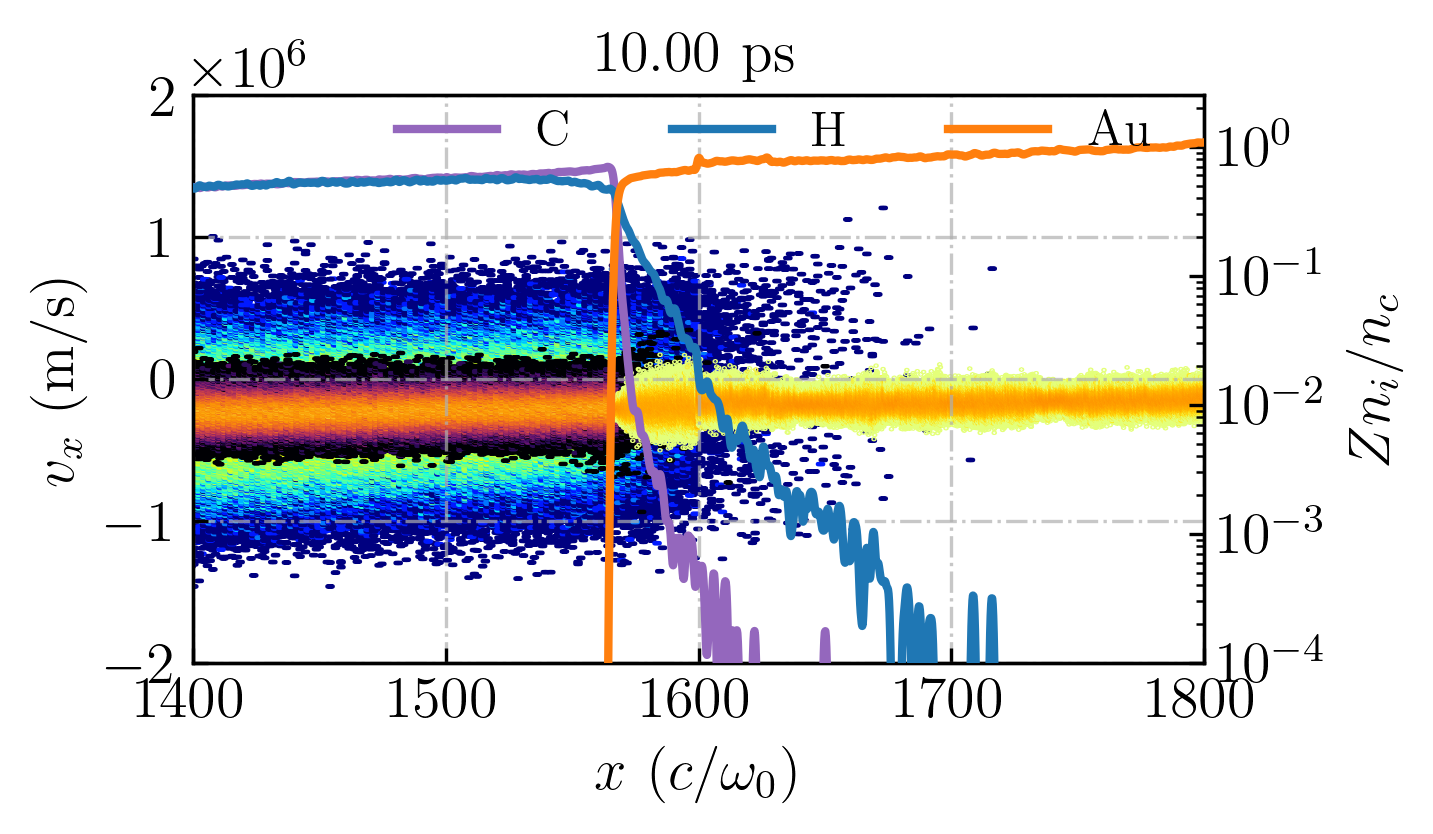}
	\includegraphics[width=0.495\textwidth]{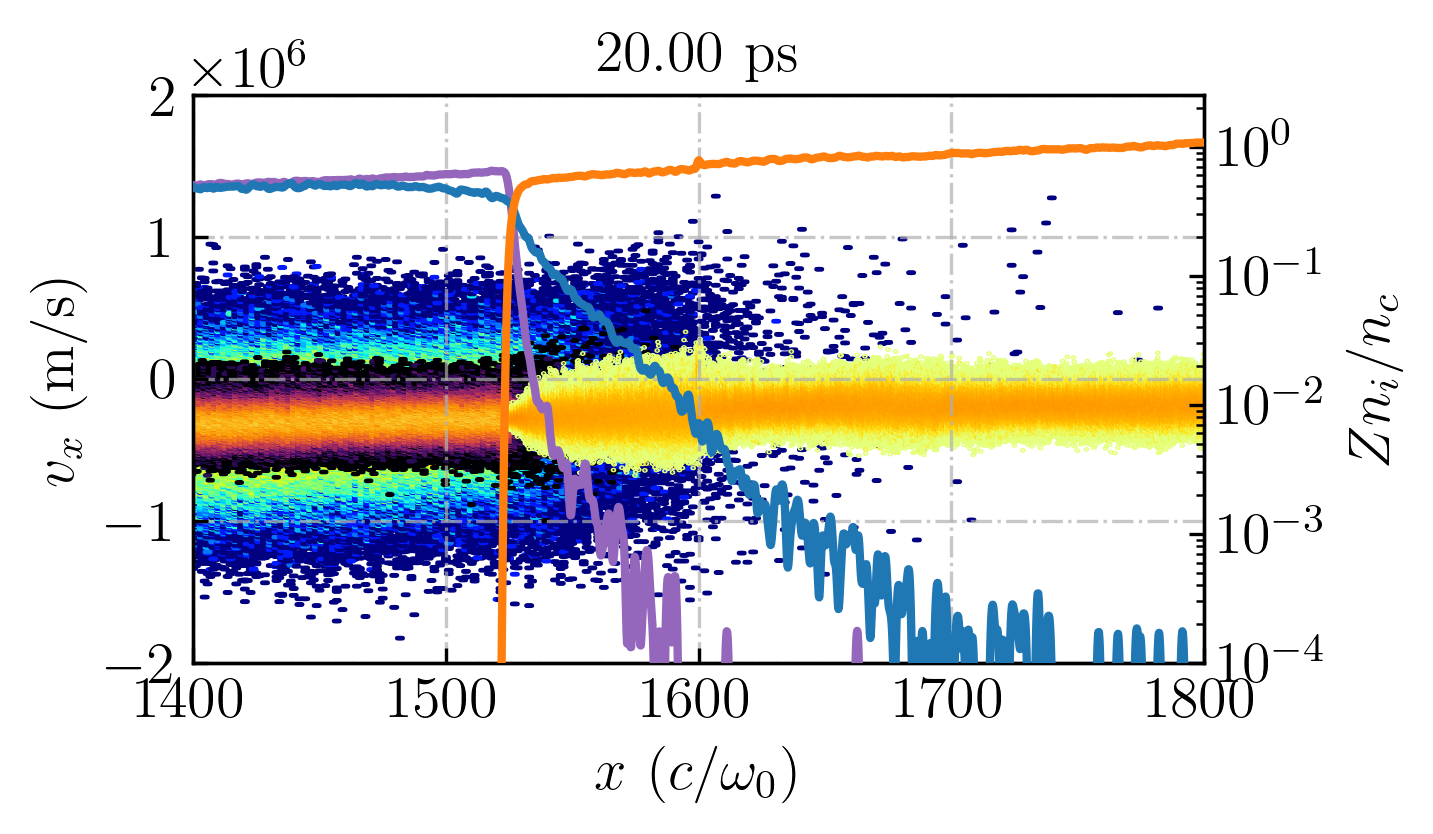}
	\caption{The phase-space distribution plots $v_x$-$x$ of the C and H ions (left half) and Au ions (right half) and the ions density of C, H and Au ions in logarithm scale for the case in a real ICF hohlraum at 10\,ps and 20\,ps.}
	\label{fig:phase_den_real}
\end{figure*}

\begin{figure}[htb]
	\centering
	\includegraphics[width=0.49\textwidth]{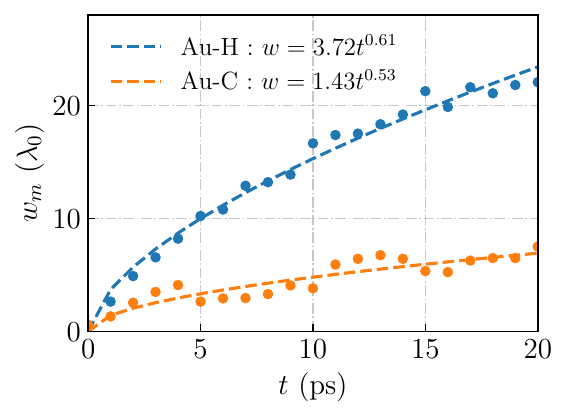}
	\caption{The mixing width of Au ions with C and H ions versus time for the case in a real ICF hohlraum.}
	\label{fig:wm_t_real}
\end{figure}

After understanding the individual influences of the incident laser, plasma flow, and density gradient on ion mixing, we simulate a case with plasma parameters extracted from RH simulations for a real ICF hohlraum, as described in section~\ref{sec:setting}. An incident laser with an intensity of $1 \times 10^{15}$ W/cm$^2$ is applied at the left boundary, as in~\cite{hao2019}.
The phase-space distribution plots of Au, C, and H ions at 10\,ps and 20\,ps are shown in figure~\ref{fig:phase_den_real}. The interface, initially at 1600$c/\omega_0$, moves to the left with the plasma flow. The C and H ions diffuse from the high-density side to the low-density side, similar to the collisional case in figure~\ref{fig:phase_den_col_wocol}, but the density profile is noisier at the diffusion front due to the combined effects of collisions and flow,
since several C and H ions are trapped by collisions with Au ions at positions far from the interface and cannot flow to the left with the background, as shown by the isolated dots on the right side of the phase-space plots. Correspondingly, the mixing widths of Au ions with C and H ions exhibit more noise as they evolve over time, as shown in figure~\ref{fig:wm_t_real}. 
The electrostatic shock wave is not observed due to the strong noise in the longitudinal electric field in this case. However, overall, the mixing speed between Au and C ions in a real hohlraum is similar to the pure diffusion case (case b) in section~\ref{sec:diffmec}, being about 5$\lambda_0$ at 20\,ps. The larger density gradient at the interface further reduces the diffusion of H ions, resulting in a shorter mixing width between Au and H ions compared to the pure diffusion case at the same time.
The mixing width between Au and C ions, $w_m \propto t^{0.53}$, is close to the classic fluid prediction of $w_m \propto \sqrt{t}$. However, for the less collisional pair of Au and H ions, $w_m \propto t^{0.61}$.

\subsection{Effects of ion mixing on SBS}

\begin{figure}[htb]
	\centering
	\includegraphics[width=0.49\textwidth]{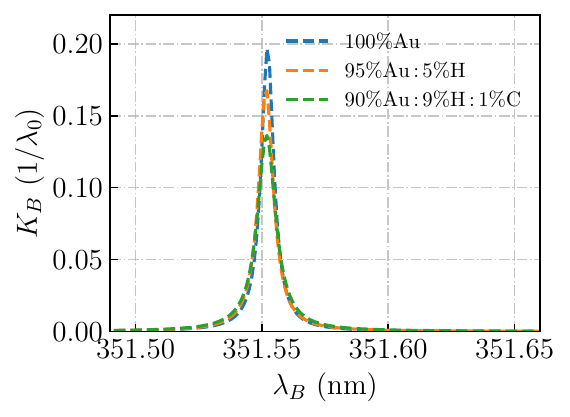}
	\caption{The spatial growth rate ($K_B$) versus scattered wavelength ($\lambda_B$) in different plasma components.}
	\label{fig:SBS}
\end{figure}

As shown in previous sections, H and C ions diffuse to the Au region as time evolves, the faster diffusion velocity of H is faster compared to C.
Thus the plasma domain can be divided to four different regions, correspondingly C$_5$H$_{12}$, Au-C-H mixing, Au-H mixing, and Au from left to right, as shown in figure~\ref{fig:phase_den_col_wocol} and~\ref{fig:phase_den_real}.
To investigate the effect of ion mixing on stimulated Brillouin scattering (SBS), we compare the spatial growth rate ($K_B$)~\cite{drake1974} in three different plasma components, namely 100\%Au, 95\%Au\,:\,5\%H and 90\%Au\,:\,9\%H\,:\,1\%C, as shown in figure~\ref{fig:SBS}.
The peak $K_B$ decreases when more H and C are added to Au plasmas, as the penetration of H and C into Au significantly enhances Landau damping in the mixing layer, which help suppress SBS~\cite{yan2019, hao2019}.

As shown in section~\ref{sec:hohlraum}, the mixing width between Au and H ions grows as $w_m = 3.72t^{0.61}$. Within a laser duration of 3,ns, the mixing width can exceed one hundred micrometers.
If the mixing layer is neglected and mistakenly considered as pure Au plasma, the SBS calculation would be artificially overestimated.
While the relationship between $w_m$ and $t$ may be more complex and may not follow a power law under ICF hohlraum conditions, we suggest that these simple fits could serve as a computationally efficient and practical approach to integrating ion-mixing processes in large-scale LPI calculation codes, such as HLIP~\cite{hao2014}.
This is particularly valuable given the high computational cost of solving multi-fluid equations in RH simulations.

\section{Summary \& Outlook}
\label{sec:conclusion}

This study presents a multi-species kinetic analysis of ion mixing at the plasma interface between hohlraum Au plasmas and filling C$_5$H$_{12}$ gases using 1D particle-in-cell simulations. Our findings reveal that ion-ion collisions significantly limit the diffusion velocity of ions, rendering Au ions sub-diffusive while C and H ions remain super-diffusive. A notable separation between C and H ions occurs due to their differing diffusion velocities. An electrostatic shock is still generated at the plasma interface even in the presence of collisions, while collisions reduce both the electric field strength and the propagation speed of the shock wave, and this electrostatic shock wave has negligible effect on the super-diffusive behavior of C and H ions.

Compared to diffusion-driven mixing at a sharp interface, an incident laser with an intensity up to $2 \times 10^{15}$\,W/cm$^2$ has a negligible effect on the ion mixing behaviors. The Au and C$_5$H$_{12}$ interface moves together with the background plasma flow, and the velocity of the electrostatic shock wave becomes the sum of the original shock velocity and the flow velocity. The mixing speed between Au and C ions slightly increases as some C ions are trapped by collisions with Au, preventing them from moving leftward with the background plasma flow. However, this phenomenon is less pronounced for the mixing between Au and H ions due to relatively lower collision rates. An inhomogeneous density profile also limits diffusion from the low-density region to the high-density region, reducing the overall mixing speed as the diffusion of C and H ions dominates the process.

Under real ICF hohlraum conditions, ion mixing behaviors are predominantly driven by diffusion, with limited impact from the above-mentioned mechanisms. The density profile near the interface is noisier due to the trapping effects of collisions and the background plasma flow. Additionally, the increasing density gradient from C$_5$H$_{12}$ to Au further reduces the mixing speed by constraining the diffusion of C and H ions.
The penetration of C and H into Au significantly enhances Landau damping in the mixing layer, which further suppresses SBS.

Our findings provide critical insights into ion mixing dynamics under ICF hohlraum conditions. We suggest that diffusion-driven mixing remains the dominant mechanism, and collisional effects play key roles in shaping the interface properties.
The simple phenomenological fits describing the mixing width evolution can serve as an efficient method for integrating ion-mixing processes in large-scale LPI codes. However, the phenomenological fits here are still relatively rough, future studies could expand on this work by developing more detailed mixing width evolution models to refine our understanding of ion mixing in ICF hohlraums.

\section*{Acknowledgments}
This work was supported by the National Natural Science Foundation of China (Grant Nos. 12275032 and 12205021) and the China Postdoctoral Science Foundation (Grant No. 2024M764280).

\section*{Data availability statement}

The data cannot be made publicly available upon publication because the cost of preparing, depositingand hosting the data would be prohibitive within the terms of this research project. The data thatsupport the findings of this study are available upon reasonable request from the authors.

\section*{References}

\bibliography{ref.bib}

\end{document}